\documentclass{article}

\usepackage[utf8]{inputenc}
\usepackage{authblk}
\usepackage{multicol}
\usepackage{graphicx}
\usepackage{amsmath}
\usepackage{amssymb}
\usepackage{siunitx}
\usepackage{authblk}
\usepackage[backend=biber, sorting=none]{biblatex}
\usepackage[a4paper, total={17cm, 26cm}]{geometry}

\title{A single-input state-switching building block harnessing internal instabilities}

\author[1]{Malte A. ten Wolde}
\author[1]{Davood Farhadi}

\affil[1]{Delft University of Technology, Department of Precision and Microsystem Engineering, Mekelweg 2, 2628 CD Delft, The Netherlands}

\date{August 2023}

\begin{document}

\maketitle

\begin{abstract}
Bistable mechanisms are prevalent across a broad spectrum of applications due to their ability to maintain two distinct stable states. Their energy consumption is predominantly confined to the process of state transitions, thereby enhancing their efficiency. However, the transition often requires two distinct inputs, implicating the requirement of multiple actuators. Here, we propose an elastic and contactless design strategy for inducing state transitions in bistable mechanisms, requiring only a single cyclic input. The strategy leverages internal information, interpreted as system state, as an extra input to make a weighted decision for transitioning to the subsequent state. We characterize the behavior using a spring-based rigid-body model, consisting of a column near bifurcation, combined with a non-linear spring connected to a bistable element that represents the information state. The results show that a nonlinear spring with a quadratic stiffness function, i.e., representing internal instability, is crucial for regulating state-switching behavior. We then demonstrate this design strategy by developing a monolithic and compliant design embodiment and experimentally evaluate its behavior. 
\end{abstract}

\section{Introduction}
Bistable mechanisms have two stable equilibrium states, separated by an energy barrier. The transition from one stable state to the other is often a snap-through action, and results in a large elastic deformation of the mechanism. These deformations consume only power when switching between the two states. These properties make bistable mechanisms suitable for various  applications across numerous fields \cite{cao_bistable_2021}, including but not limited to, (soft) robotics \cite{chen_harnessing_2018, tang_leveraging_2020}, (reprogrammable) mechanical metamaterials \cite{chen_reprogrammable_2021, zhang_design_2020}, mechanical logic structures \cite{waheed_boolean_2020, kuppens_monolithic_2021}, energy absorbers \cite{shan_multistable_2015}, in microscale electromechanical systems  (MEMS) like micro-positioners \cite{xu_design_2016}, actuators \cite{gerson_design_2012}, grippers \cite{power_bistable_2023}, optical switches \cite{chen_new_2005}, mechanical relays \cite{qiu_bulk-micromachined_2005}, and micro fluidic valves \cite{yang_thermopneumatically_2010}, and in the nanoscale, like DNA-based structures \cite{nummelin_robotic_2020, ke_structural_2018, zhou_direct_2015}.

Conventionally, the transition between the states can be initiated through an external input, such as a change in temperature, pressure, voltage, or mechanical forces, but often requires also a counter-input to reset the system. For example, stacked bistable inflatable origami modules require positive pressure (input) and negative pressure (counter-input) for multimodal deformation \cite{Melancon2022}, another bistable origami module uses compression and tension to actuate the two states and thereby creating peristaltic locomotion \cite{Bhovad2019}, and in soft media the stored elastic energy in bistable elements are used to propagate mechanical signals \cite{Raney2016}, but need to be reset manually (counter-input). These examples can be considered to be "responsive", i.e., the state of the system is a direct response to the external input - which can be considered a form of (low-level) intelligence. Intelligent systems can interact with the environment, adapt their structure to store and process information, and make autonomous decisions by tightly integrating sensing, actuation, and computation into the structure itself \cite{mcevoy_materials_2015, yasuda_mechanical_2021}. There are considered to be different levels of intelligence in systems. The next frontier of intelligent systems after responsive systems, termed "adaptive systems", will incorporate not only external inputs but also internal information, such as mechanical memory or state information \cite{walther_viewpoint_2020, kaspar_rise_2021}. Adaptive systems require less complex actuation to control a set of states, compared to responsive systems, by leaving some of the decision-making ability to the mechanism itself \cite{sitti_physical_2021}. For example, recently a pneumatically actuated soft robot was shown to walk and switch gaits to control the direction of locomotion using only a single constant source of pressurized air to actuate the robot \cite{Drotman2021}. 

Developing adaptive systems requires a building block that combines an external input with internal information when making a decision. Some research has been conducted to achieve this functionality. A well-known example is the mechanism in retractable ballpoint pens \cite{kent_writing_1965}, which is an angled tooth cam following mechanism consisting of discrete parts. When the input is pressed repeatedly, the current state of the system (ballpoint in or out) determines the next state. Generally, this functionality can be found in single-input switches, for example in MEMS devices \cite{huang_mems_2013, weight_two-position_2008, huang_novel_2020, foulds_hysteresis_2005, gao_planar_2019}. Furthermore, this functionality has been realized through different methodologies within the field of mechanical metamaterials \cite{azadpoor_mechanical_2016, kadic_3d_2019}; one such approach involves a unit cell consisting of two inward buckling beams with a carefully designed cutout, and a central 'state' beam in which its buckling direction changes when interacting with the inward buckling beams upon a cyclic input displacement \cite{Hecke2021YT}. Separate studies have demonstrated how coupled interaction between unit cells, i.e., repeated building blocks in a structure like waves in corrugated sheets \cite{bense_complex_2021} or domes in a dome-patterned sheet\cite{faber_dome-patterned_2020}, can change the response depending on their current global state. In yet another example, researchers leverage geometric frustration \cite{bertoldi_flexible_2017} to exhibit a history-dependent response, i.e., indicating that a system's past states influence its present and future behavior \cite{merrigan_disorder_2022}. 

Although there have been significant advancements in the design of these types of systems, current solutions have several limitations that hinder their performance and usability. For instance, many existing designs rely on contact-based solutions. This introduces hysteresis and displacement errors, both resulting from friction and manufacturing tolerances. In addition, they are prone to wear due to friction, which results in a loss of input information and energy. Furthermore, the predominant contact-based solutions face scalability issues, particularly when miniaturized, due to challenges like micro-stiction \cite{Spengen2003}. Additionally, unit cells with coupled interactions are often not rationally designed, making it hard to predict how these coupled interactions work when increasing the number of unit cells for more complex computations \cite{bense_complex_2021, bertoldi_flexible_2017}. Furthermore, some designs, such as those that utilize mechanically frustrated unit cells, can only compute once before having to reset manually, reducing their flexibility and versatility in real applications \cite{merrigan_disorder_2022}.

In this paper, we propose a fully elastic and contactless state-switching building block, that harnesses internal instabilities to switch between two distinct states in response to a single input. In the following sections, we will discuss the details of our proposed building block. In Section 2, we will describe the design principle behind the mechanism, including an analytical spring-based rigid-body model. Section 3 covers an analytical case study used to evaluate its performance, where the two design parameters of the nonlinear spring are studied. In Section 4 we propose a planar design embodiment and cover the numerical simulations, fabrication, and experimental validation of our prototype. Then, in Section 5, the results of the simulations and measurements are presented. Furthermore, we provide a discussion and interpretation of our findings. Section 6 will provide a discussion on opportunities and  potential future research directions. Finally, in Section 7 we present our conclusion.

\begin{figure}[t!]
    \centering
    \includegraphics[width=0.9\linewidth]{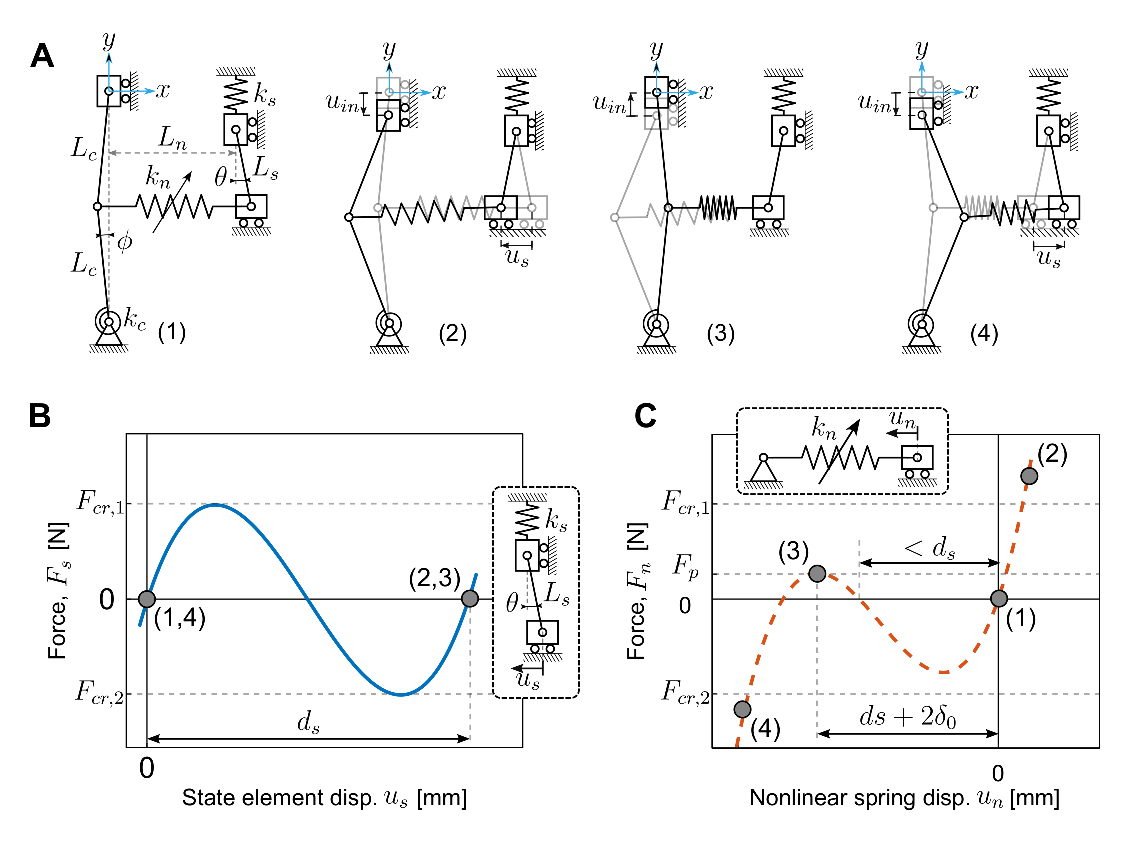}
    \caption{Design methodology for single-input state-switching building block. The building block consists of a state element, a buckling column near bifurcation, and a nonlinear spring connecting the two. (A) Four configurations of the mechanism representing the states and state-transitions: initial stable 0-state (1), state-transition from 0 to 1-state (2), stable 1-state (3), resetting state-transition (4). (B) The force-displacement characteristics of a state element. (C) One possible solution for the force-displacement characteristics of the nonlinear spring (dashed line), crossing the four key points related to the four configurations: initial conditions as fabrication (1), required tensile forces in configurations (2) and (3) to pull the buckling column to the other bifurcation path, and a compressive force in configuration (4).}
    \label{fig:states}
\end{figure}

\section{Design principle}\label{sec:principle}
To regulate state switching in bistable mechanisms, we propose a building block consisting of three elements. This includes a state element, a buckling column initially configured around bifurcation, and a connecting spring that connects the two, see Fig. \ref{fig:states}a. The building block represents two distinct states, e.g. '0-state' and '1-state', that alternate with a cyclic input displacement. A bistable mechanism, with two stable equilibrium positions, is used to represent the two states. The force-displacement characteristic of such an element is shown in Fig. \ref{fig:states}B; with state displacement $d_s$ between the two different stable states, and critical loads $F_{cr,1}$ and $F_{cr,2}$. The bistable element requires pull and push input to switch between the two states. The buckling column is configured such that it can buckle in two directions, and can convert a compression input, $u_{in}$, into a pull and push motion along the x-axis, see Fig. \ref{fig:states}A (2) and (4) respectively. Lastly, the connecting spring, shown with $k_n$, has two functions to achieve the alternating behavior. Firstly, controlling the response by reading the current state of the state element, see Fig. \ref{fig:states}A (1) and (3), and secondly, switching the state (writing) by transmitting the input force towards the state element, see Fig. \ref{fig:states}A (2) and (4). To achieve this, the spring characteristics should be highly nonlinear and meet specific design requirements. First we address the force-displacement requirements related to the four configurations displayed in Fig. \ref{fig:states}A, and then we discuss the continuous force-displacement characteristics and stiffness of the nonlinear spring. 

The connecting spring has force-displacement criteria related to each of the configurations (1) to (4), see circles labeled (1) to (4) in Fig. \ref{fig:states}C, that depend on the state element and the buckling column. In the initial configuration, the system is stable, thus the displacement $u_n=0$ and force $F_n=0$; this is represented as configuration (1). In addition, the buckling column is designed with an imperfection, i.e., a small initial angle $\phi_0$, such that the left buckling bifurcation path is preferred when an input $u_{in}$ is given. In configuration (2), when the input displacement is maximal, $u_{in}=U_{max}$, the tension in the connecting spring should exceed the critical load of $F_{cr,1}$, so that the state element snaps-through to the 1-state. In configuration (3), the connecting spring should deliver a tensile force denoted as $F_p$. This force should be equal to or greater than the force generated by the hinges of the buckling column, with stiffness $k_c$, when rotating the buckling column to $\phi=-\phi_0$. This ensures that the buckling direction of the column is in the positive x-direction. By designing the connecting spring such that the spring is still in tension when it is shortened due to state displacement $d_s$, i.e., $F_n(ds) > 0$, it can 'read' the current state and move the column through the bifurcation position 2$\delta_0$ in x-direction, with $\delta_0 = L_c\sin{\phi_0}$. Thus, in configuration (3), the following should be satisfied, 
\begin{equation} \label{eq:conFnds}
   F_n(d_s+2\delta_0) = F_p \ge 2k_c\phi_0,
\end{equation}
which is denoted as criteria $C_1$. Lastly, in configuration (4), an input displacement creates a buckling deformation along the positive x-direction, thereby generating compression in the spring. Then, when the input is maximal, $u_{in}=U_{max}$, the compression force should exceed the critical load of $F_{cr,2}$, and the state element switches back to the initial 0-state.

Through the four points (1)-(4) in Fig. \ref{fig:states}C, a continuous function, which represents the force-displacement characteristics of the connecting spring, can be plotted. One function that fits through the four points is a cubic function (e.g. the dashed line). This reveals that the connecting spring stiffness should be a quadratic form, and can be described by 
\begin{equation}\label{eq:kn}
    k_n(u_n) = \alpha (u_n-r_1)(u_n-r_2),
\end{equation}
with $u_n$ the spring displacement, and unknown variables $\alpha$, $r_1$, and $r_2$. To get the nonlinear force response ($F_n$), Eq. \ref{eq:kn} can be integrated. In the initial configuration ($u_n=0$) the force $F_n(0) = 0$, thus integrating Eq. \ref{eq:kn} yields 
\begin{equation}\label{eq:Fn}
    F_n(u_n) = \frac{\alpha u_n}{6}(2u_n^2 - 3r_1u_n - 3r_2u_n + 6r_1r_2).
\end{equation}
However, not every value for $\alpha$, $r_1$, and $r_2$ provides a viable solution. Namely, When the input is removed between config. (2) and (3), $u_{in}:U_{max}\rightarrow0$, the 1-state should remain in its stable position. This can be achieved under the condition that
\begin{equation} \label{eq:conFn23}
    F_n(u_n) > F_{cr,2} \quad \forall \, u_n \in [(2), (3)],
\end{equation}
which is denoted as criteria $C_2$. Furthermore, there are no stresses in the initial configuration, and thus configuration (1) should be the lowest energy state. The energy of the connecting spring can be determined by integrating Eq. \ref{eq:Fn}. In the initial configuration ($u_n=0$) the potential energy $E_n(0) = 0$, thus integrating Eq. \ref{eq:Fn} yields 
\begin{equation}\label{eq:conEn}
    E_n = \frac{\alpha u_n^2}{12}(u_n^2 - 2r_1u_n - 2r_2u_n + 6r_1r_2)\ge 0 \quad \forall \, u_n, 
\end{equation}
which is denoted as criteria $C_3$. 

To find viable nonlinear spring characteristics, an analysis is performed to find adequate combinations of design parameters $\alpha$, $r_1$, and $r_2$ that satisfy Eqs. \ref{eq:conFnds}, \ref{eq:conFn23}, and \ref{eq:conEn}, denoted as criteria $C_1$, $C_2$, and $C_3$, respectively. This analysis together with an analytical case study is presented in Section \ref{sec:analytical}.

\section{Analytical case study}\label{sec:analytical}

\begin{figure}[t!]
    \centering
    \includegraphics[width=\textwidth]{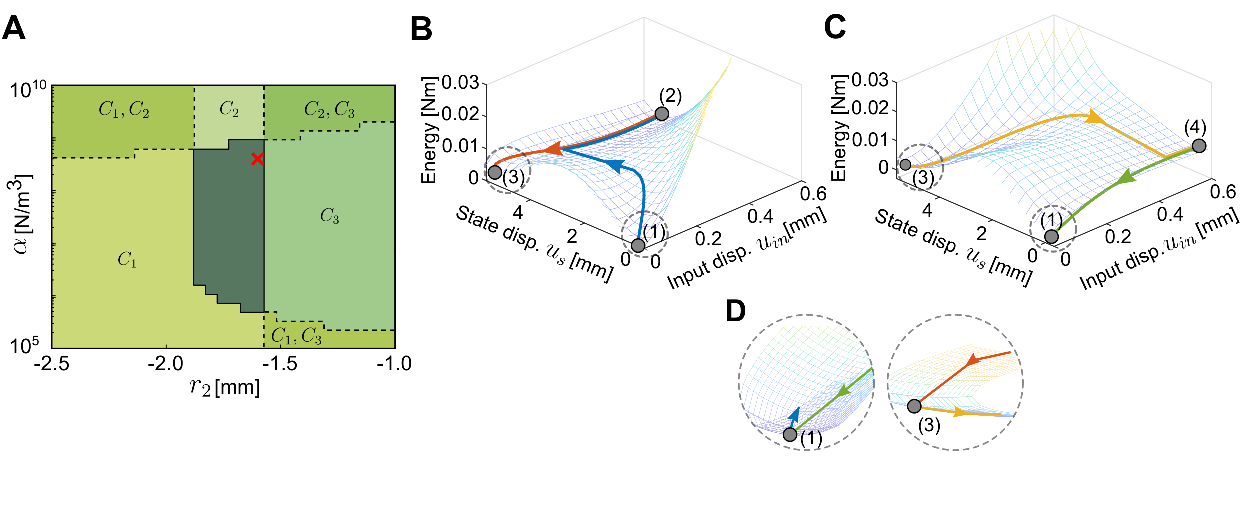}
    \caption{Analytical case study. (A) Analysis of design parameters $\alpha$ and $r_2$ for the nonlinear spring. The center dark green region represents feasible combinations, while the surrounding regions fail to meet one or more of design criteria $C_1$, $C_2$, and $C_3$. The red cross indicates the values chosen for the case study. The energy landscape of the two bifurcation paths combined with the path of minimal energy from (B) the 0-state and from (C) the 1-state. An offset of \SI{0.1}{\milli\meter} in $u_s$-direction is given for visibility of both paths. (D) The zoom-ins at (1) and (3) show the switching between bifurcation energy landscapes.}
    \label{fig:Energy}
\end{figure}

To perform a case study, first, a suitable set of connecting spring design parameters, $\alpha$, $r_1$, and $r_2$, need to be identified using the previously determined criteria $C_1$, $C_2$, and $C_3$, i.e., Eqs. \ref{eq:conFnds}, \ref{eq:conFn23}, and \ref{eq:conEn}, respectively. For the analysis, we have chosen a state element with $L_s =$ \SI{21}{\milli\meter}, $\theta_0 =$ \ang{7}, and $k_s =$ \SI{1.735e5}{\newton \per \meter}; and a buckling column with $L_c = 2.1L_s$, for sufficient geometrical advantage, and $\phi_0 =$ \ang{0.57}, as an imperfection to enforce buckling in negative x-direction. Furthermore, rotational stiffness value $k_c$, with the hinges replaced as short-length flexures, is described by 
\begin{equation}
    k_c = 4\frac{Ewt^3}{12L},
\end{equation}
with Young's modulus $E=$ \SI{1.7}{\giga\pascal}, width $w=$ \SI{7.5}{\milli\meter}, thickness $t=$ \SI{0.5}{\milli\meter}, and length $L=$ \SI{4}{\milli\meter} of the flexures in the buckling column. To satisfy Eq. \ref{eq:conFnds}, it is reasonable to position the local maxima of Eq. \ref{eq:Fn} at point (3) in Fig. \ref{fig:states}C, i.e., $r_1 = -(d_s+2\delta_0)$, such that the greatest force can be delivered; however, we note that it is not a requirement that point (3) is located at the local maximum, only that criteria $C_1$ is satisfied.

The analysis results for $\alpha$ and $r_2$ are presented in Fig. \ref{fig:Energy}A. Several distinct regions can be identified, labeled with the corresponding criteria that are not satisfied. Firstly, three vertical regions that are dependent on the energy landscape of the nonlinear spring. When $r_2<$ \SI{-1.87}{\milli\meter}, the energy landscape of the nonlinear spring shows only 1 stable point, indicating it cannot provide enough pull-in force, thus criteria $C_1$ is not satisfied. When $r_2>$ \SI{-1.57}{\milli\meter}, the energy landscape becomes non-feasible because the energy at the second stable point drops below zero, thus criteria $C_3$ is not satisfied. Furthermore, within the range of \SI{-1.87}{\milli\meter} $<r_2<$ \SI{-1.57}{\milli\meter}, three distinct horizontal regions can be identified. Firstly, when $\alpha\gtrsim$ \SI{6e8}{\newton \per \meter \cubed}, criteria $C_2$, i.e., Eq. \ref{eq:conFn23}, is not satisfied, resulting in the second state remaining unstable. Secondly, when $\alpha\lesssim$ \SI{7e5}{\newton \per \meter \cubed}, criteria $C_1$, i.e., Eq. \ref{eq:conFnds}, is not satisfied, this implies that the connecting spring does not generate enough tension to pull the buckling column to the alternate bifurcation path. A combination of design parameters in the center region fulfills all constraints and can be selected for an analytical case study.

The performance of our building block can be evaluated using the criteria on the design parameters of the connecting spring. The spring stiffness is defined as $k_n = \alpha(u_n - r_1)(u_n-r_2)$\SI{}{\newton \per \meter}, where $\alpha = $ \SI{4e8}{\newton \per \meter \cubed}, $r_1 = -(ds+2\delta_0)$ \si{\meter}, and $r_2 =$ \SI{-1.6e-3}{\meter}. These values, indicated with a red cross in Fig. \ref{fig:Energy}A, represent the physical prototype discussed in Section \ref{sec:modeling}.

The total energy of the system ($E_t$) can be calculated by summing the energy of the connecting spring ($E_n$, as described in Eq. \ref{eq:conEn}), the state element ($E_s$), and the buckling column ($E_c$). These energy components can be derived from the spring-based rigid-body model. The energy landscapes of the left and right buckling bifurcation paths are illustrated in Figs. \ref{fig:Energy}B and \ref{fig:Energy}C, respectively. Additionally, the path of minimal energy for a full input cycle of $u_{in} = 0 \rightarrow U \rightarrow 0 \rightarrow U$, where $U$ represents the maximal input displacement, is overlayed. The blue line represents the transition from 1 to 2, the red line from 2 to 3, the yellow line from 3 to 4, and the green line from 4 back to 1. This path is calculated using:
\begin{equation}
    \nabla \mathcal{L} = 0, \quad \mathcal{L}(u_{in}, u_s, \lambda) = E_t + \lambda g
\end{equation}
where $E_t$ represents the total potential energy, $g$ denotes a constraint for the input displacement $u_{in}$, and $\lambda$ is the input force required to maintain this constraint. The analysis suggests that our mechanism is capable of switching between the two states using a single input. When an input displacement is given, there is a sudden snap-through to a lower energy path, see (1) to (2) in Fig. \ref{fig:Energy}B and (3) to (4) in Fig. \ref{fig:Energy}C. This path remains stable even when the input is removed, see (2) to (3) and (4) to (1) in Figs. \ref{fig:Energy}B and \ref{fig:Energy}C, respectively. Upon reaching the new stable position, the opposite bifurcation path becomes the lowest energy path, leading the system to follow the alternate path upon the application of a new input displacement. This switching behavior, in (3) and (1), is evident in the zoom-ins, see Fig. \ref{fig:Energy}D. After completing the full cycle, the 0-state is reattained.

\begin{figure}[t!]
    \centering
    \includegraphics[width=0.85\linewidth]{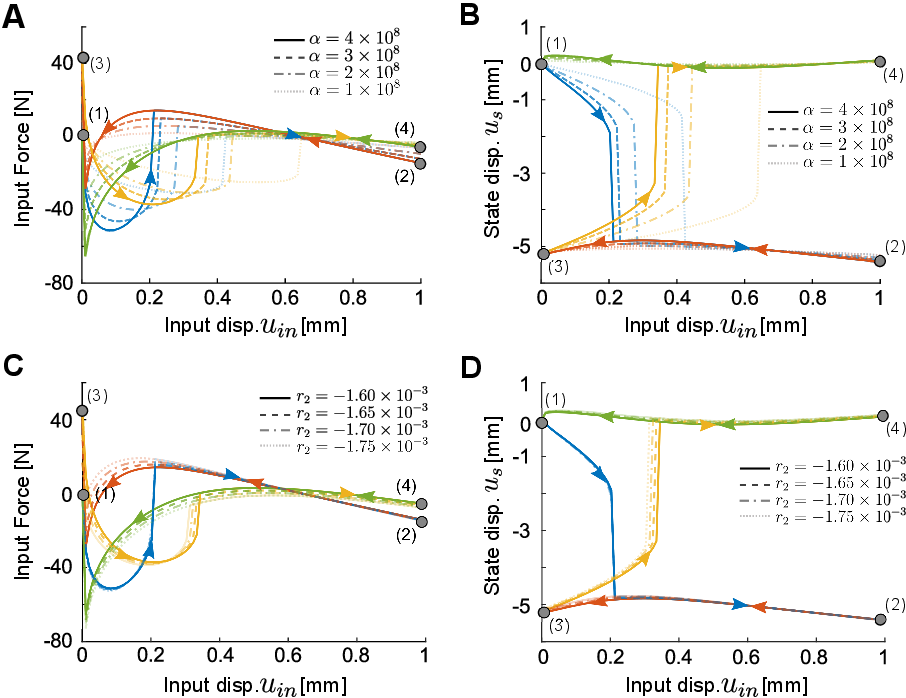}
    \caption{Performance analysis with different values of $\alpha$ and $r_2$. The blue lines show behavior from config. 1 to 2, the red lines from config. 2 to 3, the yellow lines from config. 3 to 4, and the green lines from config. 4 back to 1. The force-displacement characteristics for different values of (A) $\alpha$ and (C) $r_2$, respectively. The input-output displacement for different values of (B) $\alpha$ and (D) $r_2$, respectively.}
    \label{fig:Energy2}
\end{figure}

Using the spring-based rigid-body model, the influence of design parameters $\alpha$ and $r_2$ on the mechanism's performance can be readily evaluated, and depicted in Fig. \ref{fig:Energy2}. We analyzed four decreasing values for $\alpha$ (\num{4}, \num{3}, \num{2}, and \num{1e8}) and $r_2$ (\num{-1.60}, \num{-1.65}, \num{-1.70}, and \num{-1.75e-3}). The analysis indicates that a lower value of $\alpha$ results in a decreased input force while a larger input displacement is necessary for state switching. This can be explained by the fact that $\alpha$ is a scalar of the stiffness function, see Eq. \ref{eq:kn}. Therefore, a larger input displacement is required to overcome the critical load of the state element but with a greater transmission ratio, i.e., decreased input force. Additionally, according to Fig. \ref{fig:Energy2}C and \ref{fig:Energy2}D, the influence of $r_2$ appears to be minimal and determines if the mechanism functions as desired.

\section{Compliant design, numerical modeling, and fabrication} \label{sec:modeling}

\begin{figure}
    \centering
    \includegraphics{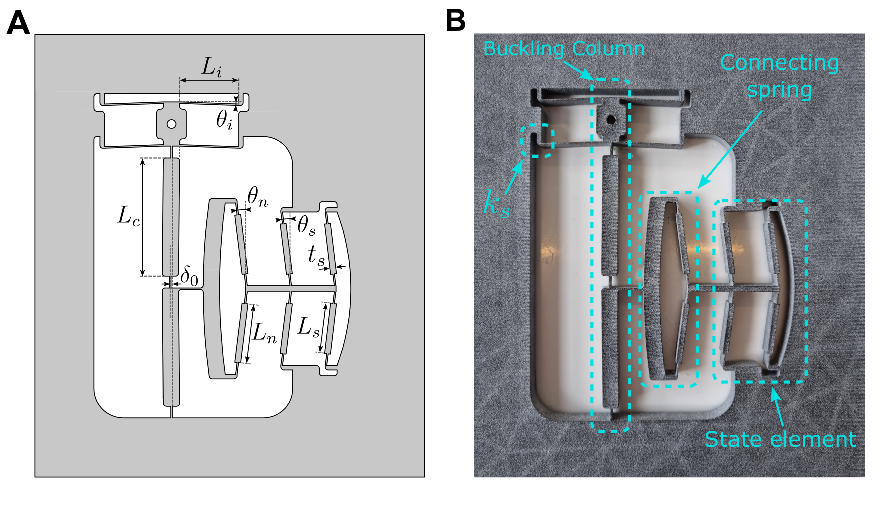}
    \caption{(A) Proposed compliant embodiment of the building block, labeled with design parameters. (B) Fabricated prototype, labeled with the three main elements. The proposed design has a out-of-plane thickness of \SI{7.5}{\milli\meter}, all flexures have a length of \SI{4}{\milli\meter} and thickness of \SI{0.5}{\milli\meter}, all compression springs ($k_s$) have a length of \SI{3.5}{\milli\meter} and thickness of \SI{1.5}{\milli\meter}, and all beams have a thickness of \SI{5.5}{\milli\meter}.}
    \label{fig:prototype} 
\end{figure}

To experimentally validate our design method, we designed an elastic, planar embodiment where all rotational hinges are replaced with small-length flexures. For the state element, we used two parallel bistable elements to provide rectilinear translation. The buckling column remains a beam with three flexures to enforce buckling. This column is suspended on a compliant shuttle to allow for rectilinear input displacement. A bistable element is a natural embodiment of the connecting spring, due to its similar cubic force-displacement behavior. This design fulfills the three criteria discussed in Section \ref{sec:principle}, and has similar values for $\alpha$, $r_1$, and $r_2$ as mentioned in Section \ref{sec:analytical}. 
Fig. \ref{fig:prototype}A shows the proposed compliant embodiment annotated with design parameters $L_s=$ \SI{17.13}{\milli\meter}, 
$t_s=$ \SI{2}{\milli\meter}, 
$\theta_s=$ \ang{7}, 
$L_n=$ \SI{20.15}{\milli\meter}, 
$\theta_n=$ \ang{7},  
$L_c=$ \SI{40}{\milli\meter}, 
$\delta_0=$ \SI{0.4}{\milli\meter},
$L_i=$ \SI{20}{\milli\meter}, and 
$\theta_i=$ \ang{2}.
Besides the annotated design parameters, the mechanism has an out-of-plane thickness of $h=$ \SI{7.5}{\milli\meter}. All short-length flexures have length $L_f=$ \SI{4}{\milli\meter} and thickness $t_f=$ \SI{0.5}{\milli\meter}, the compression springs, $k_s$, have length $L_k=$ \SI{3.5}{\milli\meter} and thickness $t_k=$ \SI{1.5}{\milli\meter}, and all other beams have a thickness $t_b=$ \SI{5.5}{\milli\meter}. 

The prototype is fabricated using 3D printing by Multi Jet Fusion (MJF) using polyamide-12 (Nylon-12). A picture of the fabricated prototype is presented in Fig. \ref{fig:prototype}B, with annotated regions that represent the state element, connecting spring, buckling column, and a region indicating one of the compression springs $k_s$.

A finite element analysis (FEA) using Ansys Parametric Design Language (APDL) was conducted to verify the mechanism's design and performance. Two-node beam elements (beam188), based on Timoshenko beam theory, with rectangular beam cross-section, were used. The mechanism's material parameters are: Young's modulus $E = $ \SI{1.7}{\giga \pascal}, density $\rho =$ \SI{1010}{\kilo\gram\per\meter\cubed}, and Poisson's ratio $\nu =$ \num{0.33} \cite{Alomarah2019}. The mechanism is anchored with fixed boundary conditions at the points where it interfaces with the frame. The design parameters were carefully selected to ensure the maximum Von Mises stress remained below the \SI{48}{\mega\pascal} limit, thereby maintaining the mechanism's structural integrity. 

Furthermore, experiments were carried out to assess the force-displacement characteristics and input-output kinematics of the mechanism. For the force-displacement measurement, a 45N force sensor (Futek LSB200 FSH03878), mounted to a precision linear stage (PI M505) is used. An input displacement of \SI{1.2}{\milli \meter} was applied to the input shuttle of the mechanism. 
Simultaneously, for the input-output displacement measurement, the displacement of the input shuttle and state element is captured using a video camera and then analyzed using image processing.

\section{Results and Discussion}

\begin{figure}
    \centering
    \includegraphics[width=\linewidth]{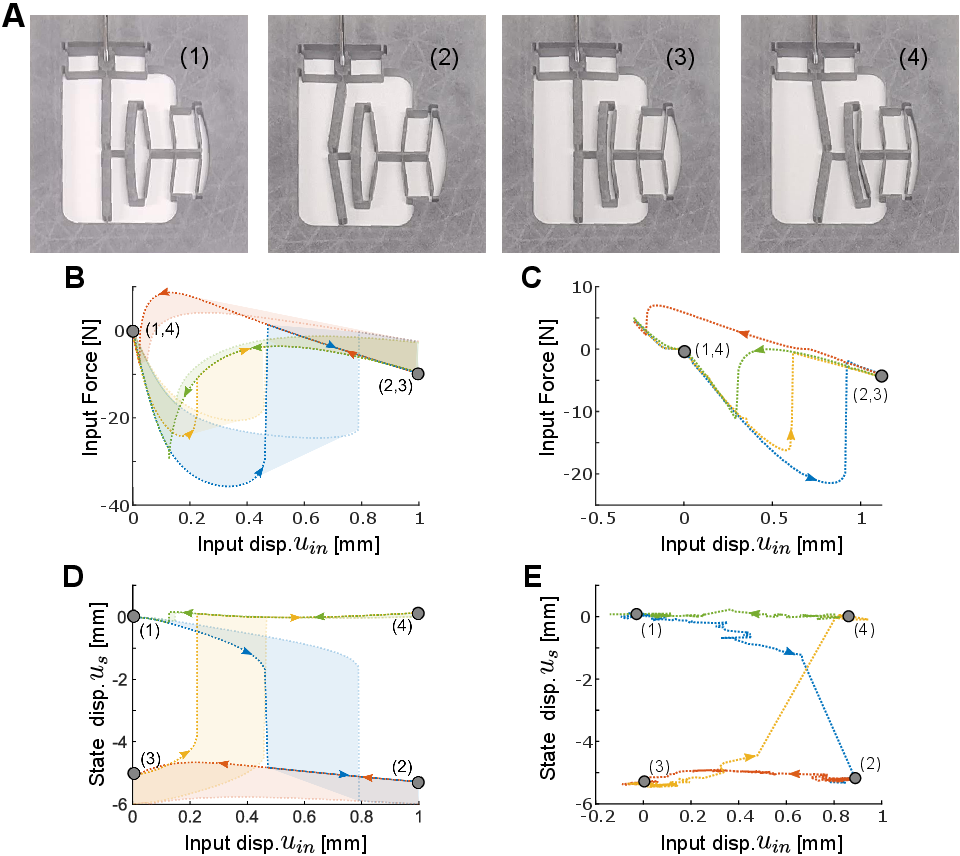}
    \caption{Results from both simulations and experimental measurements. (A) Snapshots illustrating the four configurations of the mechanism upon applying a cyclic input displacement. Force displacement characteristics from (B) simulations, and from (C) measurements. Input-output kinematics from (D) simulations, and from (E) measurements. }
    \label{fig:Results}
\end{figure}

Four distinct configurations of the developed prototype throughout a full input cycle are shown in Fig. \ref{fig:Results}A, see also video S1 (supplementary material). This sequence corresponds to an input displacement pattern of $u_{in} =$ \SI{0}{\milli \meter} $\rightarrow$ \SI{1.2}{\milli \meter} $\rightarrow$ \SI{0}{\milli \meter} $\rightarrow $ \SI{1.2}{\milli \meter}, and the configurations at each stage are denoted as (1), (2), (3), and (4) respectively. The transition from (1) to (2) illustrates the buckling of the column along with the state switching to the 1-state. In the transition from (2) to (3) the state's stability is evident, and the nonlinear spring delivers a tensile force on the buckling column. Due to the tension in the connecting spring, the buckling column follows the second bifurcation path, i.e., along the positive x-direction, upon applying an input displacement. This action prompts the transition from configuration (3) to (4), the state to reset to its original 0-state position. Lastly, in the transition from (4) to (1), upon removing the input displacement, the state remains stable and the mechanism arrives in its original configuration. 

Furthermore, the results of the force-displacement characteristics obtained by simulations and experiments are shown with dashed lines and arrows indicating the direction, see Figs. \ref{fig:Results}B and \ref{fig:Results}C, respectively. Transparent lines and areas represent fabrication inaccuracies, and will be discussed later in this section. The blue line represents the transition from 1 to 2, the red line from 2 to 3, the yellow line from 3 to 4, and the green line from 4 back to 1. Similar behavior from the analytical model is observed, see Fig. \ref{fig:Energy2}. The required input force switching to the 1-state, transition 1 to 2, is \SI{-36}{\newton} in simulation compared to \SI{-22}{\newton} in our experimental measurements. Next, when removing the input, transition 2 to 3, a maximal force of \SI{9}{\newton} in the simulation and \SI{7.5}{\newton} in the experiment is reached. Then, for switching back to 0-state, transition 3 to 4, a force of \SI{-25}{\newton} in simulation and \SI{-16.5}{\newton} in the experiment is needed. Lastly, returning to the original configuration, transition 4 to 1, the peak force is \SI{-30}{\newton} in simulation compared to \SI{-12.5}{\newton} in the experiment.

The input-output displacement results derived from both simulation and experiment are displayed in Figs. \ref{fig:Results}D and \ref{fig:Results}E, respectively. The arrows indicate the direction of the lines. The y-axis displays the state displacement, where \SI{0}{\milli\meter} displacement represents the 0-state and $\sim$ \SI{5}{\milli\meter} represents the 1-state. A sudden snap-through can be observed from 0- to 1-state and vice-versa, after the snap-through the state displacement remains stable when the input displacement is removed; this behavior is the state-switching. For transition 1 to 2, an input displacement of \SI{1.2}{\milli\meter} is applied, and the state displaces \SI{5}{\milli\meter} in simulation and \SI{5.25}{\milli\meter} in the experiments to the 1-state. The snap-through is triggered at an input displacement of \SI{0.45}{\milli\meter} in simulation and \SI{0.65}{\milli\meter} in experiments. The state then retains this position when the input is removed, transition 2 to 3, until a subsequent input is applied. The next input displacement of \SI{1.2}{\milli\meter} causes the transition from 3 to 4, upon which the state resets back to the 0-state at \SI{0}{\milli\meter} for both simulation and experiment. The snap-through occurs at an input displacement of \SI{0.23}{\milli\meter} in simulations and \SI{0.45}{\milli\meter} in the experiment.

A sensitivity analysis was performed to understand the influence of key parameters of the mechanism. Our analysis indicated that the performance is dominated by the state element, while the correct functionality of the mechanism is determined by the design criteria of the connecting spring with respect to the fabricated dimensions of the state element. 
The fabricated prototype was measured and the following design parameters were changed in the FEA:  $L_s=17.13+$\SI{0.5}{\milli\meter}, $L_k=3.5-$\SI{0.25}{\milli\meter}, $t_k=1.5+$\SI{0.25}{\milli\meter}, $\theta_s=7+$\ang{0.75}, $L_n=20.15+$\SI{0.5}{\milli\meter}, and $\theta_s=7+$\ang{1}. These values represent manufacturing inaccuracies, which are partly attributed to 3D-printer accuracy and partly because \SI{1}{\milli\meter} fillets are used in the compression springs, while the FEA used rectangular beams. Besides, the prototypes were fabricated using Material Jet Fusion (MJF); while it should be possible to fabricate flexures of \SI{0.5}{\milli\meter} using this method, a closer examination of the prototypes revealed heterogeneous material filling within the flexures. This doesn't accurately reflect the material properties used in the simulations. Therefore, additional simulations with updated dimensions were conducted using a lower Young's modulus of  \SI{1}{\giga \pascal}.
The results of the simulations are presented in Fig. \ref{fig:Results}B and \ref{fig:Results}D as transparent lines and a shaded area between the nominal values and the fabricated prototype with inaccuracies. The results indicate that small variations in the mentioned design parameters have a significant influence on the force-displacement and input-output relation. In Fig. \ref{fig:Results}B, the force-displacement characteristics, a decrease in peak force of \SI{12}{\newton} is observed in transition 1 to 2, a decrease of \SI{5}{\newton} in transition 2 to 3, a decrease of \SI{4.5}{\newton} in transition 3 to 4, and a decrease of \SI{12.5}{\newton} in transition 4 to 1. These results are similar to those measured in experiments, namely, \SI{-25}{\newton} vs \SI{-22}{\newton}, \SI{4}{\newton} vs \SI{7.5}{\newton}, \SI{-20.5}{\newton} vs \SI{-16.5}{\newton}, and \SI{-15}{\newton} vs \SI{-12.5}{\newton}, respectively. Furthermore, in Fig. \ref{fig:Results}D, the input-output displacement, the state element displaces \SI{6}{\milli\meter}, and the snap-though is triggered at \SI{0.8}{\milli\meter} in simulation vs \SI{0.65}{\milli\meter} in experiment from the 0- to the 1-state, and at \SI{0.45}{\milli\meter} in both simulation and experiment back from the 1- to the 0-state.

Further observed discrepancies between the measurements and simulations can potentially be attributed to the finite stiffness of the frame which is not considered in the FEA. Bistable structures are highly sensitive to boundary conditions. When boundary conditions are overly complaint, bistability may be lost entirely. Precautions have been taken to increase the frame's stiffness. Such as taping the prototype to a PMMA base plate, however, small changes in the boundary conditions, such as small outward displacement of the boundary conditions of the state element due to flexion of the frame, can explain some of the discrepancies. From the sensitivity analysis, it was determined that the behavior of the mechanism is dominated by the state element. An estimation of our frame stiffness is \SI{3e5}{\newton\per\meter}, which is in series connection to support stiffness $k_s$, see Figs. \ref{fig:states}A and \ref{fig:prototype}A. The support stiffness $k_s$ is estimated to be \SI{1.735e5}{\newton\per\meter}, thus the frame stiffness contributes significantly, further explaining the differences observed. 

Lastly, to actuate the mechanism, a hole of \SI{3}{\milli \meter} in diameter was implemented in the prototype to accommodate a hook attachment to provide input displacement. Due to the difference in diameter of the hole and hook, there was some hysteresis of $\sim$ \SI{0.1}{\milli\meter} in the measurement, this can be seen around \SI{0}{\milli\meter} and \SI{0.6}{\milli\meter} at \SI{0}{\newton} in Fig. \ref{fig:Results}C. This partly explains why a negative input displacement was required. 

\section{Opportunities and Outlook}
In this study, our primary focus was on the quasi-static behavior of the mechanism, with the dynamic characteristics remaining unexplored. For instance, material selection is crucial due to inherent visco-elastic behavior. As the mechanism approaches the snap-through point, the visco-elasticity can lead to relaxation behavior, thereby changing the precise snapping moment. This phenomenon becomes particularly significant during the transition from configuration 2 to configuration 3, where the forces involved are relatively low. Due to this phenomenon, besides hysteresis, we applied a small negative displacement of $u_{in}=\SI{-0.2}{\milli\meter}$ to our mechanism in the experimental study. Furthermore, to comprehensively understand the dynamic performance of the mechanism, it might be beneficial to further explore its maximum operating frequency through a multi-body dynamic model. 

When adapting our single-input state-switching mechanism for real-world applications where an output load is required, careful consideration of load placement becomes crucial. To maintain the mechanism's desired bistable behavior and eliminate unintended state changes, the output load should preferably be placed on the buckling column, e.g., at the left/right buckling point, rather than directly on the state element.

Furthermore, an interesting observation is that the input-output displacement relation of our mechanism exhibits characteristics of a frequency divider, see Fig. \ref{fig:Results}D and \ref{fig:states}E. In MEMS devices, the operation frequencies from actuators are generally high. While mechanical frequency up-conversions have been achieved before \cite{farhadi_machekposhti_compliant_2018}, down-conversion of motion frequency is rarely  reported. By concatenating multiple instances of our building block - and considering that the loads transmitted through such a system should still be carried by the input beam - we could potentially achieve higher frequency division ratios.

Lastly, while this study outlines design guidelines for the proposed mechanism, it's important to highlight the generality of our approach. To illustrate the feasibility of our design principle, we fabricated a prototype that satisfies the design criteria. However, these guidelines are not confined to this specific embodiment. For instance, our prototype embodies a rectilinear input displacement and bistable mechanism, but our design principle can adapt to other design embodiments, including rotational input displacement and other variants of bistable elements. Moreover, while our prototype is realized at the decimeter scale, the framework we present holds potential across a range of length scales, from nano to macro. Thus, our demonstrated prototype serves as a tangible representation, but our design guidelines are applicable more broadly, offering adaptation beyond the embodiment we present.

\section{Conclusion}
We have presented a fully elastic state-switching mechanism that can convert a cyclic input signal into two distinct stable states. This functionality is achieved by harnessing internal instability that guides the bifurcation path of a buckling column. By 'reading' the mechanism's current state, and 'writing' the input into the state element, it facilitates alternating switching behavior. In contrast to previous studies in which state switching has been achieved through complex contact-based interaction, we laid out a guideline for designing nonlinear springs that facilitate state switching through a fully elastic and monolithic embodiment. Although we demonstrated the theory using a centimeter scale prototype in this study, it is important to note that our approach is compatible with miniaturization, suitable for a broad spectrum of applications, from micro switches to reprogrammable metamaterials. Furthermore, the proposed methodology allows for different variations, such as changing the nature of the input motion, e.g., the type of displacement field, or adjusting the readout mode. Lastly, while this work focused on a system characterized by a single input and state element, the strategy lays the groundwork for the development of flexible mechanisms with sequencing behavior, such as sequencing between parallel sets of state elements through a single input. 

\printbibliography

\end{document}